\documentclass[letter,twocolumn]{jpsj2} 

\usepackage{times}

\usepackage{graphicx}

\title{Multipole Fluctuations in Filled Skutterudites}

\author{Takashi {\sc Hotta}}

\inst{Advanced Science Research Center,
Japan Atomic Energy Research Institute,
Tokai, Ibaraki 319-1195}

\recdate{\today}

\abst{
In order to clarify exotic multipole properties of filled skutterudites,
we evaluate multipole susceptibility for $n$=1$\sim$5,
where $n$ is the local $f$-electron number, on the basis of
a multiorbital Anderson model constructed using the $j$-$j$ coupling scheme.
For $n$=1, magnetic fluctuations dominate over low-temperature electronic
properties, while for $n$=2 and 4, electronic states are dominated
by both magnetic and quadrupole fluctuations.
For $n$=3 and 5, octupole fluctuations are found to be significant,
depending on the crystalline electric field potential.
We discuss possible relevance of the results to actual materials.
}

\kword{Multipole susceptibility, Filled skutterudites,
Multiorbital Anderson model, $j$-$j$ coupling scheme,
Numerical renormalization group method}

\begin{document}
\maketitle


Recent discovery of heavy-fermion superconductivity
in PrOs$_4$Sb$_{12}$ has triggered a rapid increase of research
activities on filled skutterudite compounds.\cite{Maple,Bauer}
This material is believed to belong to the group of
strongly correlated electron systems, but
the superconductivity cannot be simply understood in terms
of $d$-wave pairing mediated by antiferromagnetic spin fluctuations,
since the inverse of the nuclear spin-lattice relaxation time
for PrOs$_4$Sb$_{12}$ does not show the $T^3$ behavior
at low temperature $T$
characteristic of a line-node gap.\cite{Kotegawa}
In addition, PrOs$_4$Sb$_{12}$ exhibits several
exotic superconducting features such as
multiple superconducting phases,\cite{Izawa}
point-node behavior in the gap function,\cite{Izawa,Chia}
and the breaking of time-reversal symmetry
as detected by $\mu$SR experiments.\cite{Aoki1}

It is premature to settle on the mechanism of superconductivity
in PrOs$_4$Sb$_{12}$, but recently,
the possibility of exotic pairing
due to quadrupole fluctuations has been suggested
experimentally~\cite{Goto} and theoretically.\cite{Miyake}
In fact, neutron scattering experiments have revealed that
a quadrupole ordered state is induced by a magnetic field
in the vicinity of the superconducting phase of
PrOs$_4$Sb$_{12}$,\cite{Kohgi}
implying a potential role for quadrupole fluctuations
in superconducting pair formation.
Moreover, PrFe$_4$P$_{12}$ exhibits a second-order phase
transition at 6.5K,\cite{Aoki2}
but this transition is considered to be due to antiferro-
quadrupolar ordering from neutron scattering experiments
\cite{Iwasa} and theoretical efforts.\cite{Curnoe,Kiss}
It has been gradually recognized that the quadrupolar degree of
freedom plays a crucial role in filled skutterudite compounds.

From the theoretical viewpoint, multipole phenomena
in $f$-electron systems have been investigated phenomenologically
on the basis of the $LS$ coupling scheme.
This Heisenberg-like model for relevant multipole moments could
successfully explain some  of the experimental results
for filled skutterudites.
In particular, the phase diagram for PrOs$_4$Sb$_{12}$
has been reproduced \cite{Shiina1} in a quasi-quartet system
with $\Gamma_1$ ground and $\Gamma_4^{(2)}$ excited states
with a very small excitation energy \cite{CEF} for
a crystalline electric field (CEF) potential
with $T_{\rm h}$ symmetry.\cite{Takegahara}
However, the microscopic approach to electronic properties
of filled skutterudites has been very limited mainly
due to the complexity of multipole degrees of freedom.

In order to overcome such difficulties, it has been proposed
to construct a microscopic model for $f$-electron systems
based on the $j$-$j$ coupling scheme.\cite{Hotta1}
With this approach, for instance, the
microscopic origin of octupole ordering in NpO$_2$ has recently 
been established.\cite{Kubo}
Furthermore, the magnetism and superconductivity of filled skutterudite
materials have been studied from a microscopic viewpoint
on the basis of a multiorbital Anderson model.\cite{Hotta2,Hotta3}
However, the effect of multipole fluctuations has not been
investigated satisfactorily at the microscopic level,
even though their potential role has been emphasized in 
research on filled skutterudites.
In the $j$-$j$ coupling scheme, we can microscopically evaluate
physical quantities related to multipoles by treating them
as combined spin and orbital degrees of freedom.
Such a study is believed to open a door which leads to
a new stage for multipole physics of $f$-electron systems.

In this Letter, we attempt to clarify what kind of multipole
fluctuations are dominant in filled skutterudites,
by analyzing a multiorbital Anderson model based on the
$j$-$j$ coupling scheme for $n$=1$\sim$5,
where $n$ denotes local $f$-electron number.
For this purpose, we evaluate susceptibilities for fifteen kinds of
dipole, quadrupole, and octupole moments up to rank 3 using
a numerical technique.
In this way, we find that for $n$=1, magnetic fluctuations are dominant
at low temperatures, while for $n$=2 and 4, electronic states are
dominated by magnetic $and$ quadrupole fluctuations
at low temperatures.
For $n$=3 and 5, octupole fluctuations are found to appear
for some ranges of CEF parameters.
We provide some explanations and predictions on filled
skutterudites from the present results.


In general, in the $j$-$j$ coupling scheme,
a large spin-orbit interaction is first assumed,
and then, only the $j$=5/2 sextet is taken into account.
Concerning the applicability of the $j$-$j$ coupling scheme
to rare-earth compounds, readers can consult Ref.~\citen{Hotta3}.
Note that the main conduction band of filled skutterudites is
$a_{\rm u}$ with xyz symmetry,\cite{Harima1}
where the subscript ``u'' means ungerade, {\it i.e.}, odd parity for
space inversion. Then, the Anderson Hamiltonian is written as
\begin{eqnarray}
  H = \sum_{\mib{k}\sigma}
  \varepsilon_{\mib{k}} c_{\mib{k}\sigma}^{\dag} c_{\mib{k}\sigma}
  +\sum_{\mib{k}\sigma}
  (V c_{\mib{k}\sigma}^{\dag}f_{{\rm c}\sigma}+{\rm h.c.})
  +H_{\rm loc},
\end{eqnarray}
where $\varepsilon_{\mib{k}}$ is the dispersion of $a_{\rm u}$
conduction electrons with $\Gamma_7$ symmetry
in terms of the $j$-$j$ coupling scheme,
$c_{\mib{k}\sigma}$ is the annihilation
operator for conduction electrons with momentum $\mib{k}$ and
spin $\sigma$, $f_{\gamma\sigma}$ is the annihilation operator
of $f$ electrons on the impurity site with pseudospin $\sigma$
and orbital $\gamma$, $V$ is the hybridization between conduction
and $f$ electrons with $\Gamma_7$ symmetry, and $H_{\rm loc}$
is the local $f$-electron interaction term.
The orbital index $\gamma$ is introduced to distinguish three kinds
of the Kramers doublets, two $\Gamma_8$ and one $\Gamma_7$.
Here ``a'' and ``b'' denote the two $\Gamma_8$'s and ``c'' indicates
the $\Gamma_7$.
Throughout this paper, we set $V$=0.05, where
the energy unit is taken as $D$,
half of the bandwidth of the conduction band.
From band-structure calculations, the bandwidth is
estimated as 2.7 eV in PrRu$_4$P$_{12}$,\cite{Harima2}
indicating that $D$=1.35 eV.

The local $f$-electron term $H_{\rm loc}$ is given by\cite{Hotta1}
\begin{eqnarray}
  H_{\rm loc} \! &=& \! \sum_{\gamma,\sigma} B_{\gamma}
  f_{\gamma\sigma}^{\dag}f_{\gamma\sigma} +
  (1/2) \sum_{\gamma_1 \sim \gamma_4}\sum_{\sigma_1, \sigma_2}
  U^{\sigma_1,\sigma_2}_{\gamma_1, \gamma_2, \gamma_3, \gamma_4}
  \nonumber \\
  &\times& f_{\gamma_1\sigma_1}^{\dag}f_{\gamma_2\sigma_2}^{\dag}
  f_{\gamma_3\sigma_2}f_{\gamma_4\sigma_1},
\end{eqnarray}
where $B_{\gamma}$ is the CEF potential, which is already diagonalized.
It is convenient to introduce a level splitting $\Delta$
between $\Gamma_7$ and $\Gamma_8$ as
$\Delta$=$B_{\Gamma_8}$$-$$B_{\Gamma_7}$.
The Coulomb integral $U$ in the $j$-$j$ coupling scheme
is expressed in terms of three Racah parameters,
$E_0$, $E_1$, and $E_2$.\cite{Hotta1}
To set the local $f$-electron number,
we adjust the $f$-electron chemical potential for each $n$.

In order to reproduce the CEF scheme for Pr-based filled skutterudites,
we prefer positive $\Delta$.
In particular, the quasi-quartet situation with singlet ground and
triplet excited states is reproduced by taking a small positive $\Delta$.
However, since the difference between $O_{\rm h}$ and $T_{\rm h}$ is not
fully included in the $j$-$j$ coupling scheme,\cite{Hotta3}
the excited state for $n$=2 is always $\Gamma_4$ for $\Delta$$>$0,
although $\Gamma_4^{(2)}$ (a mixture of $\Gamma_4$ and $\Gamma_5$)
is the ground state in actual material.
The extent of the mixture explains the difference
between PrOs$_4$Sb$_{12}$ and PrFe$_4$P$_{12}$,\cite{Kuramoto}
but here we simply ignore such effect.
It is one of future problems to include the effect of $T_{\rm h}$
in the $j$-$j$ coupling scheme.


In order to discuss multipole properties,
we evaluate multipole susceptibilities for $f$ electrons, given by
\begin{eqnarray}
  \chi_{X}^{\Gamma \gamma} =  \frac{1}{Z} \sum_{n,m}
  \frac{e^{-E_n/T}-e^{-E_m/T}}{E_m-E_n}
  |\langle m | X_{\Gamma \gamma} | n \rangle|^2,
\end{eqnarray}
where $X_{\Gamma \gamma}$ is the multipole operator,
$X$ denoting the multipole symbol,
$\Gamma$ is the irreducible representation
with $\gamma$ to distinguish degenerate representations,
$E_n$ is the eigenenergy for the $n$-th eigenstate $|n\rangle$,
and $Z$ is the partition function given by
$Z$=$\sum_n e^{-E_n/T}$.
For $j$=5/2, we can define multipole operators
up to rank 5 in general, but we are primarily interested in
multipole properties from the $\Gamma_8$ quartet.
Thus, we concentrate on multipole moments
up to rank 3.\cite{Shiina2}
Note that in the $j$-$j$ coupling scheme,
physical quantities for multi-$f$-electron systems can be evaluated
from one-particle operators on the basis of quantum-field theory,
as we have performed for $d$-electron systems.
We have the clear advantage that it is not necessary to redefine
$X_{\Gamma \gamma}$ depending on the ground-state multiplet,
when we change local $f$-electron number.


Now let us show explicit forms for multipole operators
following Ref.~\citen{Shiina2}.
As for dipole moments with $\Gamma_{\rm 4u}$ symmetry,
we express the operators as
\begin{equation}
 J_{{\rm 4u}x}=J_x,~J_{{\rm 4u}y}=J_y,~J_{{\rm 4u}z}=J_z,
\end{equation}
where $J_x$, $J_y$, and $J_{z}$ are three angular momentum operators
for $j$=5/2, respectively.
Concerning quadrupole moments, they are classified into
$\Gamma_{\rm 3g}$ and $\Gamma_{\rm 5g}$, where g denotes gerade.
We express the $\Gamma_{\rm 3g}$ quadrupole operators as
\begin{equation}
\begin{array}{rcl}
 &&O_{{\rm 3g}u} = (2J_z^2-J_x^2-J_y^2)/2, \\
 &&O_{{\rm 3g}v} = \sqrt{3}(J_x^2-J_y^2)/2.
\end{array}
\end{equation}
For the $\Gamma_{\rm 5g}$ quadrupole, we have the three operators 
\begin{equation}
\begin{array}{rcl}
 &&O_{{\rm 5g}\xi} = \sqrt{3} \, \overline{J_yJ_z}/2,\\
 &&O_{{\rm 5g}\eta} = \sqrt{3} \, \overline{J_zJ_x}/2,\\
 &&O_{{\rm 5g}\zeta} = \sqrt{3} \, \overline{J_xJ_y}/2,
\end{array}
\end{equation}
where the bar denotes the operation of taking all possible
permutations in terms of cartesian components.

Finally, regarding octupole moments, there are three types as
$\Gamma_{\rm 2u}$, $\Gamma_{\rm 4u}$, and
$\Gamma_{\rm 5u}$.
We express the $\Gamma_{\rm 2u}$ octupole as
\begin{equation}
  T_{\rm 2u}=\sqrt{15} \, \overline{J_xJ_yJ_z}/6.
\end{equation}
For the $\Gamma_{\rm 4u}$ octupole, we express the operators as
\begin{equation}
\begin{array}{rcl}
 &&T_{{\rm 4u}x}=(2J_x^3-\overline{J_xJ_y^2}-\overline{J_xJ_z^2})/2,\\
 &&T_{{\rm 4u}y}=(2J_y^3-\overline{J_yJ_z^2}-\overline{J_yJ_x^2})/2,\\
 &&T_{{\rm 4u}z}=(2J_z^3-\overline{J_zJ_x^2}-\overline{J_zJ_y^2})/2,
\end{array}
\end{equation}
while the $\Gamma_{\rm 5u}$ octupole operators are given by
\begin{equation}
\begin{array}{rcl}
 &&T_{{\rm 5u}x}=\sqrt{15}(\overline{J_xJ_y^2}-\overline{J_xJ_z^2})/6,\\
 &&T_{{\rm 5u}y}=\sqrt{15}(\overline{J_yJ_z^2}-\overline{J_yJ_x^2})/6,\\
 &&T_{{\rm 5u}z}=\sqrt{15}(\overline{J_zJ_x^2}-\overline{J_zJ_y^2})/6.
\end{array}
\end{equation}


For the evaluation of multipole susceptibility,
we resort to the numerical renormalization group (NRG) method,
\cite{NRG} in which momentum space is logarithmically
discretized to include efficiently the conduction electrons
near the Fermi energy.
In actual calculations, we introduce a cut-off $\Lambda$ for
the logarithmic discretization of the conduction band.
Due to the limitation of computer resources,
we keep $m$ low-energy states.
In this paper, we set $\Lambda$=5 and $m$=3000.
Note that the temperature $T$ is defined as
$T$=$\Lambda^{-(N-1)/2}$ in the NRG calculation,
where $N$ is the number of the renormalization step.


In Figs.~1(a)-(e), we show the temperature dependence of
$T\chi_X^{\Gamma\gamma}$ for $E_0$=5, $E_1$=2, and $E_2$=0.5
with $\Delta$=$10^{-5}$.
We note that such a small value of $\Delta$ is taken
to reproduce the CEF level scheme for
Pr-based filled skutterudites.\cite{Hotta2,Hotta3}
Note also that we show one of $\chi_X^{\Gamma\gamma}$'s in each
irreducible representation,
since susceptibilities in the same irreducible representation
take the same values owing to cubic symmetry.
First of all, in the degenerate region with very small $\Delta$,
electron-hole symmetry approximately holds within the $j$=5/2 sextet.
Namely, the results for $n$=1 and $n$=2 are quite similar to those
for $n$=5 and $n$=4, respectively, as observed in Figs.~1(a)-(e).
For $n$=1 and 5, at low temperatures, we observe that
$\Gamma_{\rm 4u}$ magnetic octupole fluctuations are predominant.
Note that the fluctuations of the $\Gamma_{\rm 4u}$ dipole also survive,
since it belongs to the same symmetry as the $\Gamma_{\rm 4u}$ octupole.
Also, for $n$=3, there remain magnetic fluctuations of the $\Gamma_{\rm 4u}$
dipole and octupole.

For $n$=2 and $4$, on the other hand, we find significant fluctuations
for the $\Gamma_{\rm 5g}$ and $\Gamma_{\rm 3g}$ quadrupoles,
in addition to those of the $\Gamma_{\rm 4u}$ dipole and octupole.
Let us explain the results for the case of $n$=2.
As shown in Fig.~1(f), for $\Delta$$>$0, the local $f$-electron
ground state is a $\Gamma_1$ singlet composed of two $\Gamma_7$ electrons.
The first excited state is a $\Gamma_4$ triplet formed by $\Gamma_7$
and $\Gamma_8$ electrons, while the second excited state is a $\Gamma_5$
including a couple of $\Gamma_8$ electrons.
Even if the $\Gamma_1$ singlet is the ground state,
there exists a $\Gamma_4$ triplet state
with very small excitation energy.
This $\Gamma_4$ triplet state carries a quadrupole moment.
Thus, significant quadrupole fluctuations remain for $n$=2.
The same explanation works also for $n$=4 due to the electron-hole
symmetry for very small $\Delta$.

\begin{figure}[t]
\begin{center}
\includegraphics[width=8.5truecm]{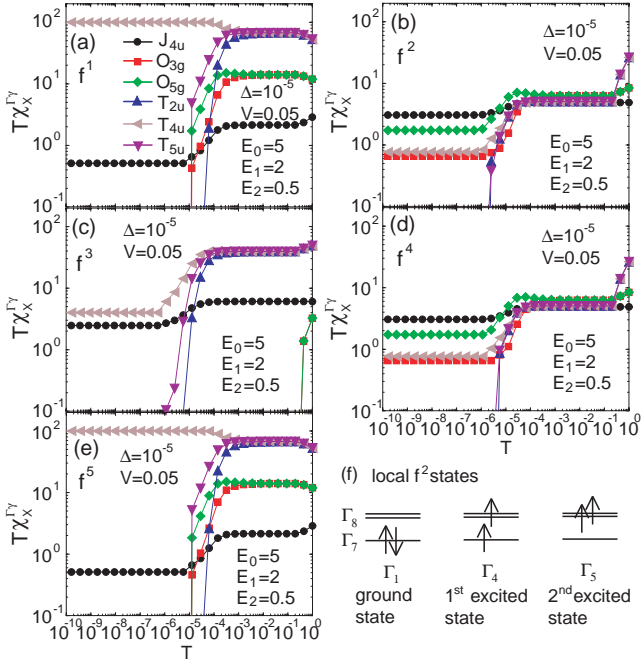}
\caption{Multipole susceptibilities vs. temperature
at $\Delta$=$10^{-5}$ for (a) $n$=1, (b) $n$=2, (c) $n$=3, (d) $n$=4,
and (e) $n$=5.
(f) Local $f^2$ states in the $j$-$j$ coupling scheme.}
\end{center}
\end{figure}

Now we consider the case with larger $\Delta$,
since the CEF potential is easily changed due to the substitution
of rare-earth ion R, transition metal ion T, and pnictogen X
for the filled skutterudite RT$_4$X$_{12}$.
In Figs.~2 (a)-(e), we show the temperature dependence of
$T\chi_X^{\Gamma\gamma}$ for $\Delta$=0.1.
For $n$=1, the low-temperature behavior of multipole
susceptibilities is almost unchanged from those
for $\Delta$=$10^{-5}$.
Since one $f$ electron is always located in a $\Gamma_7$ orbital,
$\Gamma_8$ is irrelevant for the case of $n$=1.
Since the Kondo temperature is very low for $V$=0.05,
the local moment of the $\Gamma_7$ electron still persists
in the present temperature range, but it should be screened
eventually for extremely low $T$.
For $n$=2, the $\Gamma_1$ singlet ground state is well separated
from the $\Gamma_4$ triplet excited state and thus, we do not
observe any fluctuations at low temperatures.

For $n$=3 and $5$, we note another electron-hole relation
in the $\Gamma_8$ quartet for very large $\Delta$,
as is easily understood from Fig.~2(f).
In fact, the main results in Figs.~3(c) and (e) look similar
to each other.
In particular, we find that $\Gamma_{\rm 2u}$ octupolar fluctuations
are predominant at low temperatures for $n$=3 and 5.
Note that $\Gamma_{\rm 4u}$ and $\Gamma_{\rm 5u}$ octupolar
fluctuations also become significant for $n$=5.
Enhancement of octupolar fluctuations is a characteristic point
for $n$=3 and 5 in the region of large $\Delta$.
Finally, for $n$=4 with large $\Delta$,
since the $\Gamma_7$ levels are fully occupied,
a couple of electrons in the $\Gamma_8$ quartet form a local $\Gamma_5$
triplet, leading to dominant magnetic fluctuations.

\begin{figure}[t]
\begin{center}
\includegraphics[width=8.5truecm]{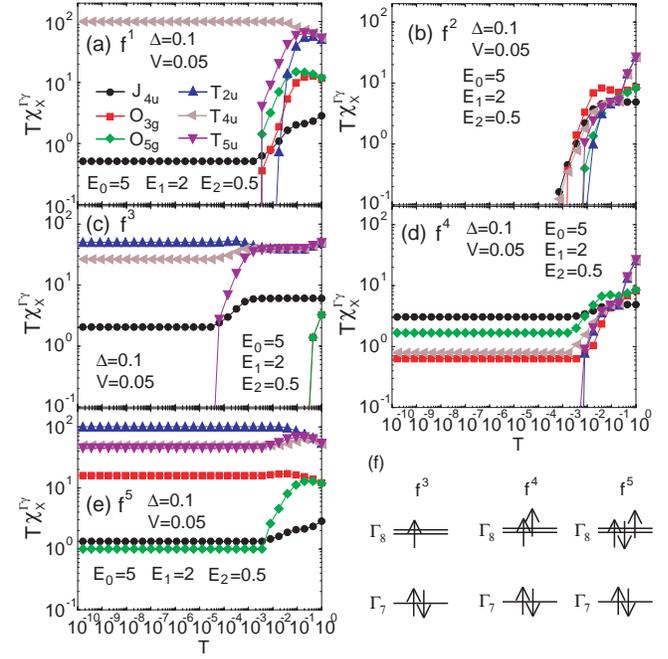}
\caption{Multipole susceptibilities vs. temperature at $\Delta$=0.1
for (a) $n$=1, (b) $n$=2, (c) $n$=3, (d) $n$=4, and (e) $n$=5.
(f) Local $f$-electron ground states for $n$=3, 4, and 5 with large
$\Delta$ in the $j$-$j$ coupling scheme.}
\end{center}
\end{figure}

In order to see the change of multipole fluctuations,
it is convenient to show the Curie constant for multipole
susceptibility,
defined as $C_{X}^{\Gamma \gamma}$=$T\chi_{X}^{\Gamma \gamma}$
at $T$=$10^{-10}$.
In Figs.~3(a)-(e). we plot $C_{X}^{\Gamma \gamma}$ vs. $\Delta$
for each $n$.
In Fig.~3(f), we summarize the multipole properties.
For $n$=1, as mentioned above, the multipole properties
do not depend on $\Delta$.
Note again that for $T$$\ll$$10^{-10}$, the local moment of the$\Gamma_7$
electron should be screened eventually.
For $n$=2, in the region of $\Delta$$<$5$\times$$10^{-5}$,
both magnetic and quadrupole fluctuations are dominant,
while for $\Delta$$>$5$\times$$10^{-5}$,
multipole fluctuations are all ``dead''.
For $n$=3, in the region of $\Delta$$<$0.02,
$\Gamma_{\rm 4u}$ magnetic octupole and dipole fluctuations
are dominant, even though the absolute values are small.
On the other hand, for $\Delta$$>$0.02,
$\Gamma_{\rm 2u}$ octupole fluctuations turn to be dominant.
For $n$=4, $\Gamma_{\rm 4u}$ dipolar fluctuations are always predominant,
but quadrupolar fluctuations gradually die out
as $\Delta$ increases.
Finally, for $n$=5, in the region of $\Delta$$<$2$\times$$10^{-4}$,
$\Gamma_{\rm 4u}$ magnetic octupole fluctuations are predominant,
while for $\Delta$$>$2$\times$$10^{-4}$,
fluctuations of the $\Gamma_{\rm 2u}$ octupole are dominant
and those of $\Gamma_{\rm 4u}$ and $\Gamma_{\rm 5u}$
are subordinate, similar to the results for $n$=3
with very large $\Delta$, due to 
electron-hole symmetry in the $\Gamma_8$ subspace.

\begin{figure}[t]
\begin{center}
\includegraphics[width=8.5truecm]{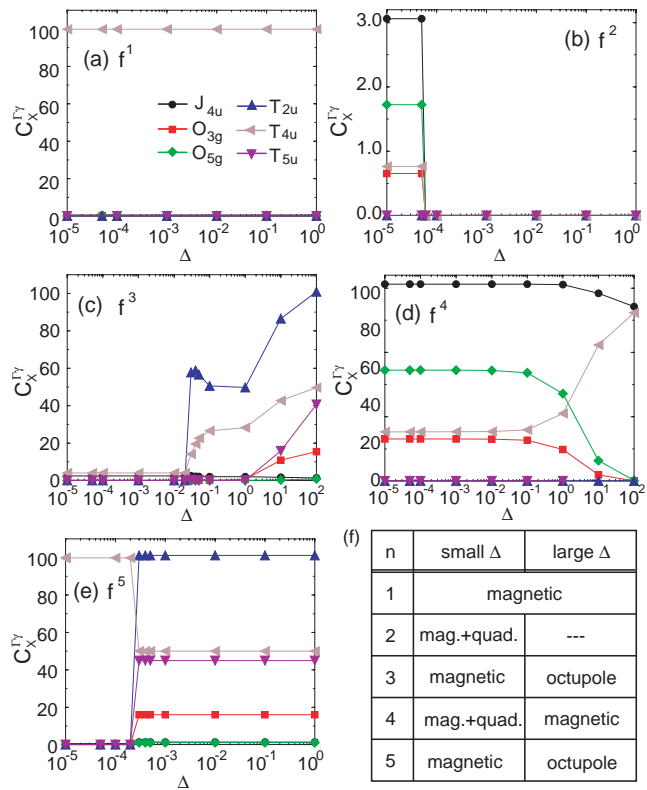}
\caption{Curie constant for each multipole susceptibility
as a function of $\Delta$ for
(a) $n$=1, (b) $n$=2, (c) $n$=3, (d) $n$=4, and (e) $n$=5.
(f) Summary for the types of relevant multipole.
Here ``mag.'' denotes magnetic, while ``quad.'' indicates quadrupole.
The boundary between small and large $\Delta$ depends on $n$,
but we simply ignore such a quantitative point in this table.}
\end{center}
\end{figure}

Next we discuss relevance of our results to RT$_4$X$_{12}$.
Although the actual system is not exactly described by
the impurity model, the multipole properties
are deduced from the dominant fluctuations.
The case of $n$=1 would seem to correspond to the case of R=Ce,
but in filled skutterudites,
the cerium ion is considered to be tetravalent,
rather than trivalent.
Thus, it may be difficult to claim a direct relation
between the $n$=1 results and CeT$_4$X$_{12}$.
For $n$=2, on the other hand, we expect a clear correspondence
to the case of R=Pr, in which
we confirm significant magnetic and quadrupole
fluctuations at low temperatures.
The range of $\Delta$ with magnetic and quadrupole
fluctuations is very narrow, which may be related to
the experimental fact that the appearance of
exotic superconductivity is very limited
in Pr-based filled skutterudites.

The cases of $n$=3 and 5 correspond to R=Nd and Sm, respectively.
For both cases, in the region of small $\Delta$,
the predominance of magnetic fluctuations suggests that
magnetic ordering occurs in actual systems,
although we have not determined whether it is ferro or antiferro.
If it is possible to increase $\Delta$ experimentally
for Nd- and Sm-based filled skutterudites,
we expect exotic electronic properties to result from
octupolar fluctuations, including $\Gamma_{\rm 2u}$
and $\Gamma_{\rm 5u}$.
In fact, for SmRu$_4$P$_{12}$, signs of octupolar ordering have been
suggested by elastic constant measurements \cite{Yoshizawa} and
$\mu$SR experiments.\cite{Hachitani}
When we turn our attention to actinides, quite recently,
NpFe$_4$P$_{12}$ has been successfully synthesized.\cite{Aoki}
Since the neptunium ion is considered to be tetravalent,
Np-based filled skutterudites correspond to $n$=3.
Thus, it may be natural to consider that
the low-temperature properties of Np-based filled skutterudites
are dominated by magnetic fluctuations.

It is difficult to find actual filled skutterudite materials
corresponding to $n$=4, but Pu-based filled skutterudites
may be good candidates if they can be synthesized,
since actinide ions are tetravalent in the filled skutterudite
structure.
For $n$=4 with relatively small $\Delta$,
based on an approximate electron-hole relation
in the $j$-$j$ coupling scheme,
one expects to find electronic properties
similar to those of Pr-based filled skutterudites.
In particular, exotic superconductivity may also occur
in Pu-based filled skutterudites with relatively high
transition temperatures,
since hybridization is large in actinide compounds
owing to the difference in the nature of $4f$ and $5f$ electrons.
This point should be carefully discussed, when Pu-based filled
skutterudites are synthesized at some future point.

Two comments are in order.
(i) In this paper, we have simply ignored the effect of the rattling
motion of rare-earth ion,\cite{Goto}
but it is interesting to see whether magnetic and
quadrupolar fluctuations are enhanced or not due to the effect
of rattling in Pr-based filled skutterudites.
This point can be investigated based on an extended model
incorporating further electron-phonon interaction terms
into $H_{\rm loc}$. This is among possible future developments.
(ii) We have suggested $\Gamma_{\rm 2u}$ for the dominant type of
octupole moment for $n$=5, but in the $j$-$j$ coupling scheme,
the effect of $T_{\rm h}$ symmetry is not fully included. 
Since $\Gamma_{\rm 4u}$ and $\Gamma_{\rm 5u}$ octupole moments should
also occur in actual filled skutterudites, we cannot conclude which is
the winner among octupole moments for R=Sm from the present results.
To resolve this point, it is necessary to analyze a model
including seven $f$ orbitals, as has been done in Ref.~\citen{Hotta3}.
This is another future problem.


In summary, we have analyzed the multiorbital Anderson model
to evaluate multipolar susceptibilities.
It has been found that for $n$=1, magnetic fluctuations are always
significant at low temperatures.
For $n$=2, electronic states are dominated by magnetic and
quadrupolar fluctuations at low temperatures,
while multipolar fluctuations are dead for large $\Delta$.
For $n$=3 and 5, at small $\Delta$ magnetic fluctuations are dominant,
while for large $\Delta$, $\Gamma_{\rm 2u}$ octupolar fluctuations are
found to occur.
For $n$=4 at small $\Delta$, the situation is quite similar to
the case of $n$=2, but for increasing $\Delta$, quadrupolar fluctuations
gradually die out.


The author thanks D. Aoki, H. Fukazawa, Y. Haga,
H. Harima, K. Kaneko, Y. Kohori, K. Kubo, T. D. Matsuda,
N. Metoki, H. Onishi, and M. Yoshizawa for discussions.
He also thanks R. E. Walstedt for critical reading of the manuscript.
This work is supported by Japan Society for the Promotion of Science
and by the Ministry of Education, Culture, Sports, Science,
and Technology of Japan.
The computation has been done using the facilities
of the Supercomputer Center of Institute for Solid State Physics,
University of Tokyo.


\end{document}